\documentclass[aps,prb,twocolumn,superscriptaddress,showpacs]{revtex4-1}

\usepackage{graphics,color}

\begin{document}

\title{
Calculated Curie temperatures for rare-earth permanent magnets:\\
{\it ab initio} inspection on localized magnetic moments in $d$-electron ferromagnetism
}

\author{Munehisa~Matsumoto}
\altaffiliation{
    Present address: Institute of Materials Structure Science,
    High Energy Accelerator Research Organization, 1-1 Oho, Tsukuba, Ibaraki 305-0801, Japan}
\author{Hisazumi~Akai}
\affiliation{
  Institute for Solid State Physics (ISSP), University of Tokyo,
  Kashiwa-no-ha 5-1-5, Kashiwa 277-8581, JAPAN}
\affiliation{
  Elements Strategy Initiative Center for Magnetic Materials (ESICMM),
  National Institute for Materials Science (NIMS),
  Sengen 1-2-1, Tsukuba 305-0047, JAPAN}

\date{\today}

\begin{abstract}
We present a data set of calculated Curie temperatures
for the main-phase compounds of rare-earth permanent magnets.
We employ {\it ab initio} electronic structure calculations
for the itinerant ferromagnetism and an effective spin model for the finite-temperature magnetism.
Curie temperatures are derived on the basis of a classical Heisenberg model
mapped via Liechtenstein's formula for atomic-pair-wise
exchange couplings. Relative trends with respect to the species of rare-earth elements
in calculated Curie temperatures for R$_{2}$Fe$_{14}$B are in agreement with experimental trends.
Quantitative comparison between calculation and experimental data found in the literature
points to an effective range of the exchange couplings
imposing a limit on the validity range of the effective spin model.
\end{abstract}


%
%
%

\maketitle

\section{Introduction}
\label{sec::intro}

Global concern for sustainable energy supply in terms
of environmental friendliness, preservation of natural resources
puts rare-earth permanent magnets (REPM's) as one of the most
critically important materials
for the future of human society. Today's commercial
REPM's are made of Fe and rare-earth elements like Nd, 
and issues associated with the usage of REPM at high temperatures
have posed a renewed interest in the magnetism
of Fe-based materials at finite temperatures.
In today's commercial champion magnet the main-phase
compound Nd$_2$Fe$_{14}$B~\cite{sagawa_1984,croat_1984,rmp_1991}
comes with a relatively low Curie temperature
585~K, almost only a half of the Curie temperature of elemental Fe at
1043~K. Considering the high-temperature range of practical usage
of traction motors of cars up to 450~K, it has been in high demand
to find a way to supplement the high-temperature
performance of Nd-Fe-B magnets or to design a new champion magnet desirably
with higher Curie temperatures with a good structure stability.

\begin{figure}
\begin{center}
\scalebox{0.8}{
\includegraphics{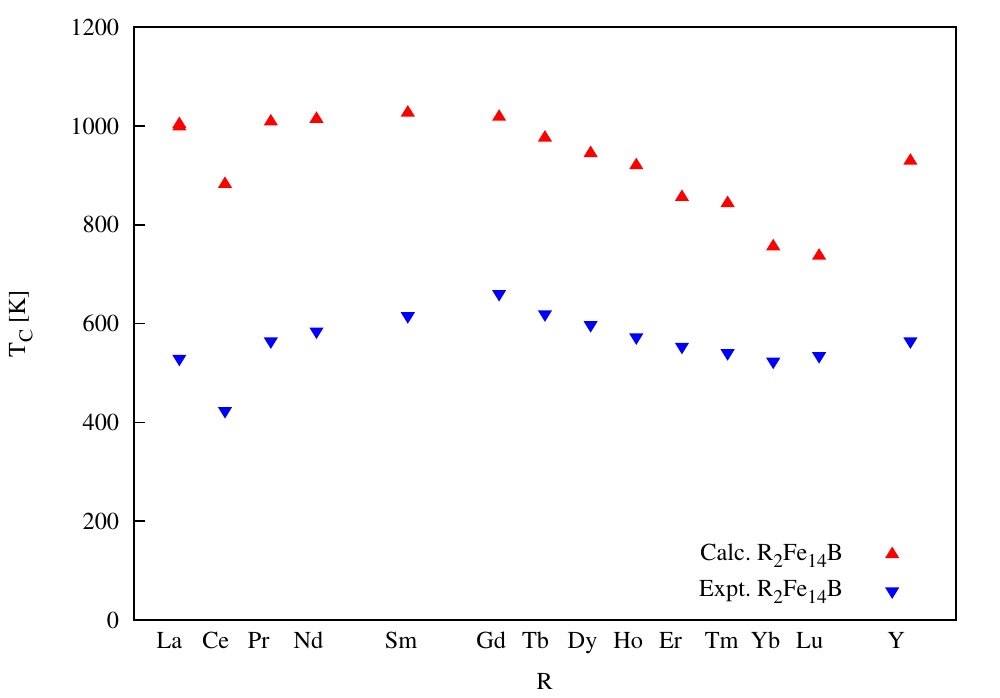}
}
\end{center}
\caption{\label{fig::Akai_T_Curie} (Color online) Calculated Curie temperature $T_{\rm C}$
for the champion magnet compound, Nd$_{2}$Fe$_{14}$B, and isostructural materials R$_2$Fe$_{14}$B
(R=rare earth) compared with experimental data.
The input lattice information is taken from past experimental data~\cite{rmp_1991}.
}
\end{figure}

The question posed to solid state physics may
boil down to finding an optimal material in a space spanned by
a) magnetization, b) magnetic anisotropy, and c) Curie temperature,
all of which backed-up by d) structure stability~\cite{JOM_2015}.
In the present study we focus on Curie temperature and magnetization
which come from the leading-order energy scales in the intrinsic electronic structure
of those $4f$-$3d$ intermetallic compounds.
The Curie temperature comes from the dominating
$3d$-$3d$ direct exchange couplings in the order of 10~meV for each exchange path,
which is multiplied by the coordination number in the lattice that is typically from 10 to up
to around 20 in the immediate neighborhood of the magnetic atom~\cite{mm_2016}.
Even a tiny amount of coupling in the long-range exchange path could contribute a lot
considering the huge coordination number~\cite{toga_2016}. However, consideration
of too long-range exchange path involves a problem in resolving the localized magnetic moment
contribution and the obvious itinerant contribution to the intrinsic magnetism. It is an important
problem to understand how well we can make {\it ab initio} prediction for those important properties
a) - d) both for fundamental theory of metallic magnetism and for practical demands in the past
and upcoming decades. The leading-order answer to the materials family including the
champion magnet compound is presented in Fig.~\ref{fig::Akai_T_Curie}.

Clarification of the validity range of a spin model for itinerant magnetism
would possibly lead to several possibilities for an optimal compound for permanent magnet utility
in the middle of subtle interplay between localized
and delocalized nature in the $d$-electron state.
Here a possible lesson
may be found in the case of an exceptionally high
Curie temperature for a Ce-based ferromagnet CeRh$_3$B$_2$ at $120~\mbox{K}$~\cite{cerh3b2_synthesis,cerh3b2_Tc}
that is even higher than that of the Gd-counterpart at $90~\mbox{K}$.
This problem has been known
for almost the same duration of history as for Nd$_2$Fe$_{14}$B since the early 1980's
and the particular dual nature showing both of localization and delocalization
as probed by X-ray absorption spectroscopy~\cite{imada}
in $4f$-electron state of Ce has been discussed. Electronic states
on the crossover between localized and delocalized electronic states are
most fertile both in terms of fundamental solid state physics and practical functionalities
and we inspect the problem for $d$-electron ferromagnetism with some additional resolution
of the particular exceptional contribution from Ce in Ce$_2$Fe$_{14}$B.

The present paper is organized as follows.
In the next section, our computational methods are described.
Main results are shown in Sec.~\ref{sec::results}.
Discussions on the difference between Fe-rich limit
and Co-rich limit and lattice structure variants are given in Sec.~\ref{sec::disc}.
The final section is devoted for conclusions and outlook.

\section{Methods}
\label{sec::methods}

For the champion magnet compound
Nd$_2$Fe$_{14}$B and related ferromagnets,
we address the ferromagnetism and the expected Curie temperature on top of it
via {\it ab initio} electronic structure calculations
on the basis of Korringa-Kohn-Rostoker Green's function method
as implemented in ``AkaiKKR''~\cite{AkaiKKR}.
Our calculations are based on
local density approximation (LDA),
taking the exchange correlation function following
Vosko, Wilk and Nusair~\cite{vosko_1980}.
Presented below are data on the basis of atomic-sphere approximation (ASA).
Muffin-tin approximation basically yields the same trends as has been given by ASA
for the present purpose to address magnetization and exchange couplings.
No further parameters are involved in our first principles calculations:
we will see that plain LDA works for $4f$-$3d$ intermetallics
which may be compared to the description of the ferromagnetic ground state of bcc-Fe on the basis of LDA.

Electronic state of rare-earth elements are treated within open-core approximation
in order to inspect the trends among
the exchange couplings between $3d$-electrons that
dominate the Curie temperature
in the leading order~\cite{skomski_1998, mm_2016}.
Trivalent state is assumed for rare-earth elements except for Ce.
Ce is set to be in tetravalent state in Ce-Fe/Co intermetallics, due to the exchange splitting
of conduction bands spanning the energy width of a few eV's
while the $4f$-electron level in Ce compounds, if localized,
sits at a position as shallow as only up to 2~eV below the Fermi level~\cite{mm_2010}.
Thus it is not very likely for the $4f$-electron in Ce to be kept in the localized level when it
is put in the exchange-split conduction band made of $3d$-electrons coming from Fe or Co.

Recently, construction of spin models for finite-temperature magnetism
of Nd$_2$Fe$_{14}$B has been attempted
by several groups~\cite{toga_2016,nishino_2017,toga_2018,westmoreland_2018,tsuchiura_2018,gong_prb_2019,gong_prmater_2019,gong_2020}.
We employ most simplified form of them to give an overview of the trend in the calculated Curie temperatures.
Non-relativistic calculations are employed
which should be sufficiently precise
to extract the trends in magnetization and exchange couplings.
For a converged ferromagnetic state of each target material,
inter-atomic exchange couplings are determined following
Liechtenstein {\it et al}.~\cite{liechtenstein_1987}.
An effective spin Hamiltonian
which goes like the following
\(
{\cal H}=-\sum_{\left<i,j\right>}2J_{i,j}S_{i}S_{j}\mathbf{e}_i\cdot\mathbf{e}_{j}
\)
is used to survey the trends in the Curie temperature on the basis of mean-field approximation (MFA).
Here the site index $i$ specifies an atom on which localized magnetic moment of the magnitude
$S_{i}$ and the direction as pointed by the unit vector $\mathbf{e}_i$ is assumed. Summation over
the exchange path between sites $i$ and $j$ runs only once on each path $\left<i,j\right>$.
The range of the exchange couplings is incorporated without any cutoff
and all of the exchange couplings in the metallic ferromagnetism
are directly plugged into the mean-field approximation on the basis of KKR Green's function method.
Only $3d$-electrons from Fe or Co and $5d$-electrons from rare earth elements
are considered in the spin Hamiltonian, on the basis of the open-core approximation for rare-earth elements.
Systematic error originating in MFA can be eliminated by a numerically exact solution
of the spin Hamiltonian via Monte Carlo methods~\cite{jaswal, mm_2016, toga_2016}
while more fundamental error between
theory and experiment can come in from the discrepancy between itinerant ferromagnetism
and localized spin model. This is what we wish to address through the inspection
of the trends between the MFA data on the basis of {\it ab initio} electronic structure and the
experimental data.

\begin{figure}
\begin{center}
\scalebox{0.8}{
\includegraphics{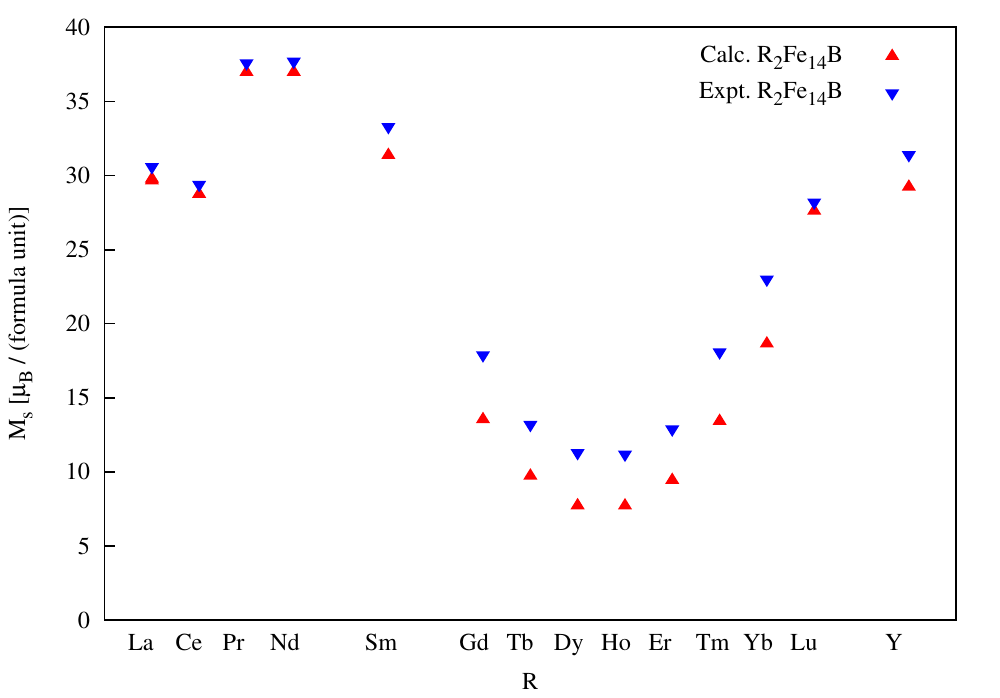}
}
\end{center}
\caption{\label{fig::Akai_mag} (Color online)
Calculated magnetization for R$_2$Fe$_{14}$B (R=rare earth)
compared with experimental data~\cite{rmp_1991}. In our results
based on open-core treatment for rare earth, the contribution from $4f$-electrons
has been manually added by $\pm g_J\sqrt{J(J+1)}$ where the sign $+$/$-$ is used for
light/heavy rare earth, respectively,
except for R=La, Ce, Lu, and Y where no $4f$ electron contribution was assumed. Here $J$ is
the total magnetic moment for each rare earth and $g_J$ is Land\'{e} $g$-factor.}
\end{figure}
\section{Results}
\label{sec::results}
\begin{figure}
\begin{center}
\scalebox{0.8}{
\includegraphics{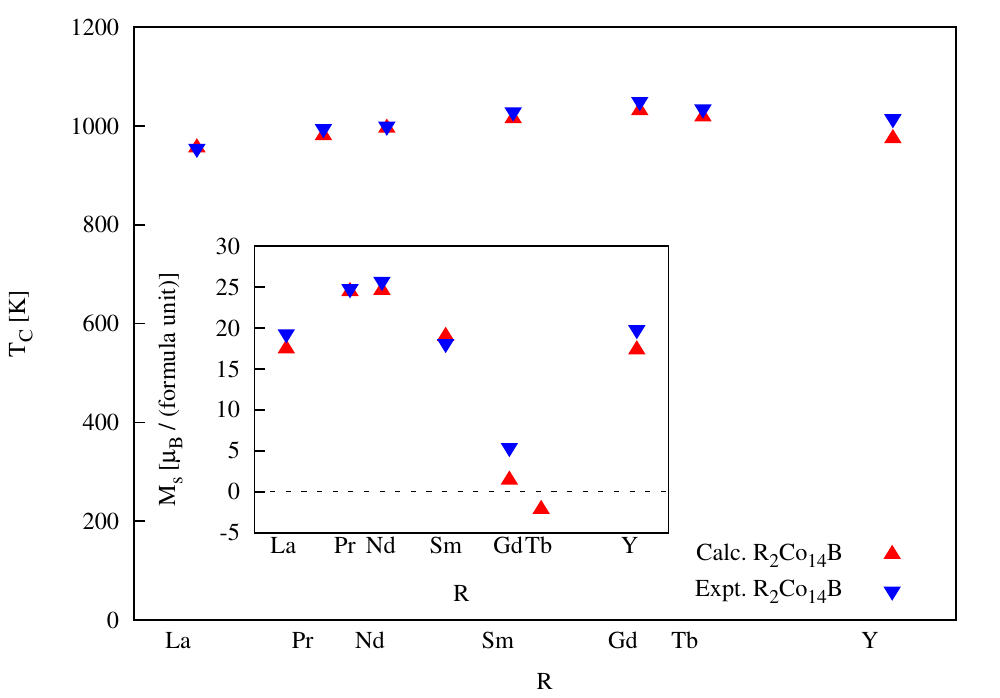}
}
\end{center}
\caption{\label{fig::Akai_Tc_and_M_for_Co} (Color online)
Calculated Curie temperature for R$_2$Co$_{14}$B (R=La, Pr, Nd, Sm, Gd, Tb, and Y).
In the inset, magnetization of R$_2$Co$_{14}$B calculated in the same way as is done
in Fig.~\ref{fig::Akai_mag} is shown. For the input to our calculations,
experimental lattice constants are taken from Ref.~\onlinecite{rmp_1991}.
Experimental Curie temperature and magnetization are also
found in Ref.~\onlinecite{rmp_1991}.}
\end{figure}

Calculated magnetization for R$_2$Fe$_{14}$B is shown in Fig.~\ref{fig::Akai_mag}.
Taken together with the results shown in Fig.~\ref{fig::Akai_T_Curie},
experimental trends of the intrinsic ferromagnetism in the 2:14:1 material family
with respect to the species of rare-earth (RE) elements are reasonably well reproduced
on the basis of our electronic structure calculations
involving only $d$-electrons. Intrinsic ferromagnetism concerning the exchange couplings
has been well described up to the leading-order, on the basis of the precision
for the total magnetization where the $4f$-electron contribution can be restored on top of
the results from open-core calculations. Results for Co-analogues
are shown in Fig.~\ref{fig::Akai_Tc_and_M_for_Co}. Overall parallel trends between
theory and experiments again hold but with even a better quantitative precision.

\begin{figure}
\begin{center}
\scalebox{0.8}{
\includegraphics{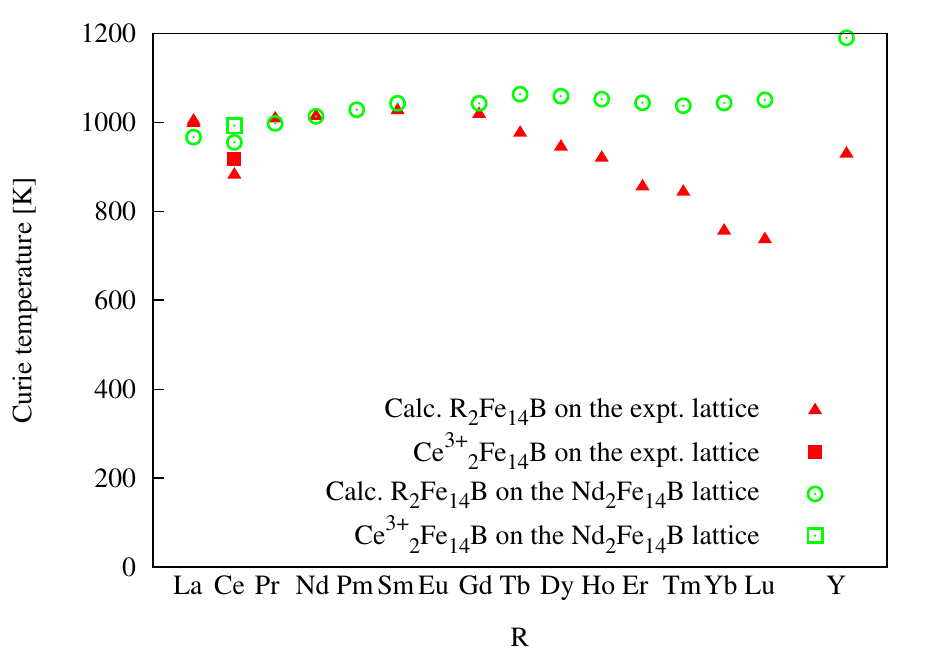}
}
\end{center}
\caption{\label{fig::lattice_diff} (Color online)
Comparison of calculated Curie temperatures for R$_2$Fe$_{14}$B on the fixed lattice
of Nd$_{2}$Fe$_{14}$B compared to the calculated data with the experimental lattice constants.
For a further comparison, results with Ce$^{3+}$ has been included. See the text for details.}
\end{figure}
\begin{figure}
\begin{center}
\scalebox{0.8}{
\includegraphics{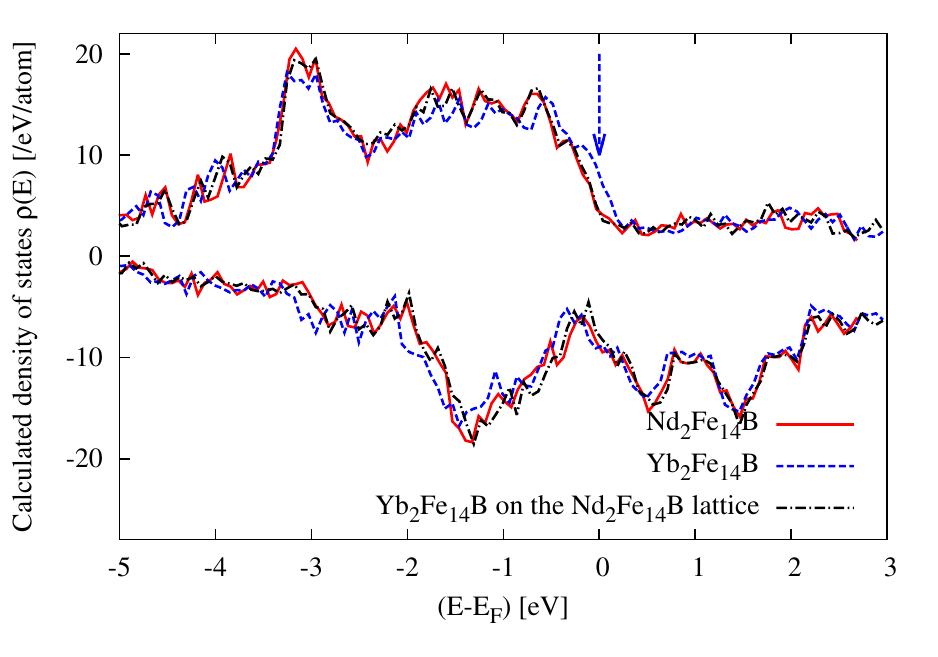}
}
\end{center}
\caption{\label{fig::dos_diff} (Color online)
Comparison of calculated density of states for Nd$_2$Fe$_{14}$B,
Yb$_{2}$Fe$_{14}$B with the experimental lattice constants
and a fictitious Yb$_{2}$Fe$_{14}$B on the Nd$_{2}$Fe$_{14}$B. The arrow
in the figure points to the majority-spin density of states for the experimental-lattice
Yb$_{2}$Fe$_{14}$B indicating the weak-ferromagnetism.
}
\end{figure}

The decreasing trend in the Curie temperature of R$_2$Fe$_{14}$B
with R being the heavy rare-earth elements
can be attributed to lanthanide contraction which seems to trigger the lattice
contraction as we overlap another calculated Curie temperatures with the lattice
fixed to be that of Nd$_{2}$Fe$_{14}$B and dependence on the chemical composition $R$
is monitored. The results for this line of reasoning is presented in Fig.~\ref{fig::lattice_diff}.
The particular drop of the Curie temperature for Ce$_2$Fe$_{14}$B in the overall trend
in R$_{2}$Fe$_{14}$B can be attributed to the exceptional tetra-valent state and the corresponding
lattice shrinkage. As seen in Fig.~\ref{fig::lattice_diff}, fictitiously restoring trivalent state for Ce
brings up the Curie temperature and putting Ce$^{3+}_2$Fe$_{14}$B
on the same fixed lattice of Nd$_2$Fe$_{14}$B brings about the smooth trend in calculated Curie temperatures,
in strong contrast to the parallel trends of calculated Curie temperatures
and experimental ones in Fig.~\ref{fig::Akai_T_Curie}.
Inspecting the difference in the density of states (DOS) for R$_2$Fe$_{14}$B on the fixed lattice
and that on the experimental lattice with the lanthanide contraction, say for R=Yb$^{3+}$
as shown in Fig.~\ref{fig::dos_diff}, we see that the lanthanide contraction
has made the $d$-electron magnetism to weak ferromagnetism in contrast to the strong ferromagnetism
found for Nd$_{2}$Fe$_{14}$B~\cite{kitagawa_and_asari_2010}.

\section{Discussions}
\label{sec::disc}

\subsection{Inspection on the range of exchange interaction}
\begin{figure}
\begin{center}
\scalebox{0.8}{
\includegraphics{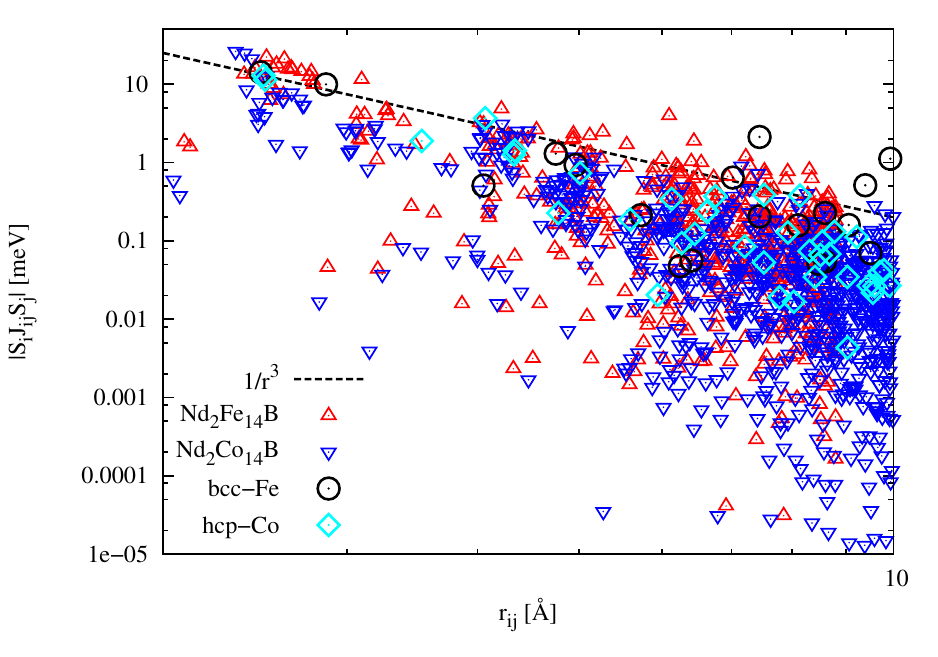}
}
\end{center}
\caption{\label{fig::Akai_Jij} (Color online)
Calculated exchange couplings for Nd$_2$T$_{14}$B
(T=Fe and Co) are shown with the absolute values plotted
as a function of the inter-atomic distances.
The analogous data for bcc-Fe and hcp-Co are overlapped as a reference.}
\end{figure}

\begin{figure}
\begin{center}
\scalebox{0.8}{
\includegraphics{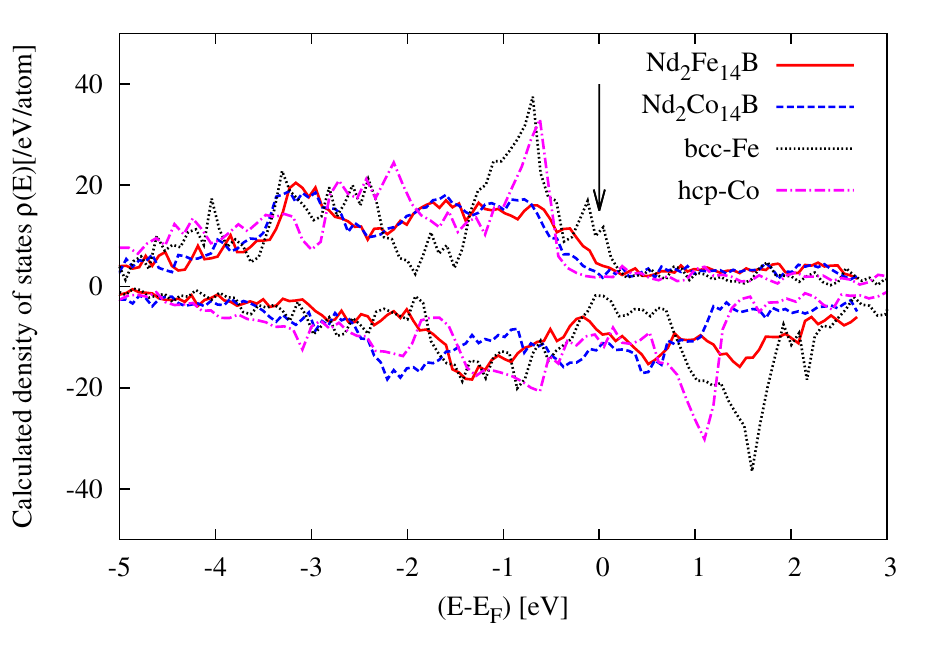}
}
\end{center}
\caption{\label{fig::Akai_dos} (Color online)
Calculated density of states for Nd$_2$T$_{14}$B
(T=Fe and Co) overlapped with bcc-Fe and hcp-Co. The arrow on the Fermi level
points to the majority-spin states which show the weak ferromagnetism.}
\end{figure}

We have seen that the effective Heisenberg model can describe
the trends of magnetization and Curie temperature for R$_2$T$_{14}$B with respect to R
while the quantitative levels achieved for T=Fe and T=Co look quite different.
To see how a spin model can be acceptable for the itinerant ferromagnetism,
we come back to the reference elemental case studies
with bcc-Fe ($a=5.417$~a.u.) and hcp-Co ($a=4.7377$~a.u. and $c/a=1.623$).
Calculated Curie temperature is 1200~K for bcc-Fe and 1411~K for hcp-Co,
in a reasonable agreement with experimental data 1043~K for bcc-Fe
and 1394~K for hcp-Co on the basis of MFA. The exchange couplings in the spin Hamiltonian
as calculated following Liechtenstein {\it et al}'s approach~\cite{liechtenstein_1987}
is shown in Fig.~\ref{fig::Akai_Jij} together with the corresponding data
for Nd$_{2}$Fe$_{14}$B and Nd$_{2}$Co$_{14}$B. The range of
Fe-Fe exchange couplings is longer than Co-Co exchange couplings.
Long-range couplings reflect the inherently itinerant electronic states~\cite{PRL_2016}.
The trend from weak ferromagnetism to strong ferromagnetism goes like
bcc-Fe, Nd$_{2}$Fe$_{14}$B, Nd$_{2}$Co$_{14}$B and hcp-Co as can be seen in the calculated
DOS shown in Fig.~\ref{fig::Akai_dos}.
Long-range exchange couplings comes from the metallic nature in the electronic state.
The overall trend is that the less metallic the better description has been achieved
on the basis of the spin model.
The description of bcc-Fe could have been worse
if the minority spin band had more 
DOS around the Fermi level. An apparent
reasonable description of bcc-Fe on the basis of the simplified spin model might depend
on the particular density of states where
the diminishing DOS in the minority-spin band renders the system less metallic.
This may compensate for the weak ferromagnetic nature with significant
DOS in the majority-spin band, out of which yielded ferromagnetism may not have been
very suited for the description based on localized spins.

Strictly speaking, a spin-only model for itinerant magnetism is invalid
by construction for quantitative studies
while discussions focusing on relative trends in a restricted range of
the chemical composition and the lattice structure
may well be justified~\cite{mm_2016,toga_2016}
up to the leading order.
Besides, anomalous temperature dependence of the lattice constants~\cite{andreev_1995}
observed in R$_2$Fe$_{14}$B has been completely dropped from our present description
of Curie temperature. Proper incorporation of finite-temperature lattice dynamics
should improve the calculated Curie temperature that is dominated by itinerant $d$-electron magnetism.
Incorporation of the lattice effect in the description
of finite-temperature magnetism
on elemental Fe and Co has been done recently
and relevance
of lattice effects especially for bcc-Fe has been shown~\cite{ruban_2018}.
Common physics may well be at work in Nd$_2$Fe$_{14}$B.

\subsection{Crystal structure trends}
\label{sec::struc}

For Nd$_{2}$Fe$_{17}$ our calculated Curie temperature
is $T_{\rm Curie}=670$~K. Compared with the experimental data 326~K~\cite{isnard_et_al_1992},
the systematic overestimate is coming
by a factor of 2.1 which is on a par with what we got for Nd$_{2}$Fe$_{14}$B
where $T^{\rm calc.}_{\rm Curie}/T^{\rm expt.}_{\rm Curie}\simeq 1.7$.
For Sm$_2$Co$_{17}$, our calculated Curie temperature
is $T_{\rm Curie}=1173$~K which compares well with the experimental number for 1200K.
Calculated Curie temperature is strongly influenced by the chemical composition:
in the overall virtual parameter space spanned by the chemical composition and crystal structure,
the composition of $3d$-metals plays a dominant role. In a ``cross section'' specified
by a fixed chemical composition the relative trends with respect to RE elements can be satisfactorily
addressed.

\section{Conclusions and outlook}
\label{sec::conc}

We have shown that
the leading-order
trends of the Curie temperature in the champion magnet compound family
with respect to the species of rare-earth elements can be described
from first principles focusing on the dominating $d$-electrons.
Care must be taken in addressing
the trend of calculated Curie temperatures in Fe-Co alloy
on the basis of the effective spin model.
The validity range of the spin-only description quantitatively differs
between the all-Fe limit and all-Co limit. The strong ferromagnetism
region in the Co-rich side characterized by the robust localized magnetic
moment seems to be better suited for the spin model for a quantitative description.
Within a subspace of fixed chemical composition,
the relative trends in the intrinsic magnetism can be satisfactorily
addressed on the basis of the effective spin model.

Having clarified the validity range of the spin model written
in terms of localized degrees of freedom for the intrinsically
delocalized ferromagnetism of $4f$-$3d$ intermetallics, it is hoped
that several possibilities for an optimal ferromagnetism for permanent magnet utility
can be identified in the middle of
subtle interplay between localized
and delocalized nature in the $d$-electron state.

\begin{acknowledgments}
This work was supported by Toyota~Motor~Corporation
and the Elements Strategy Initiative Center for Magnetic Materials
(ESICMM) under the outsourcing project of the Ministry of Education,
Culture, Sports, Science and Technology (MEXT), Japan.
Helpful discussions with collaborators in related projects
are gratefully acknowledged.
\end{acknowledgments}

\end{document}